\documentclass[
    ,final            
    ,numberedheadings 
  ,bibliocites
  ]
  {aipproc}

\layoutstyle{6x9}

\usepackage{natbib}

\begin{document}

\title{CDM Substructure in Gravitational Lenses:  Tests and Results}

\author{C.S. Kochanek}{
  address={Center for Astrophysics, MS-51, Cambridge MA 02138}
}

\author{N. Dalal\footnote{Hubble Fellow}~}{
  address={School of Natural Sciences, Institute for Advanced Study,
    Princeton NJ 08540}, 
}

\begin{abstract}
We use a simple statistical test to show that the anomalous flux ratios observed 
in gravitational lenses are created by gravitational perturbations from 
substructure rather than propagation effects in the interstellar medium 
or incomplete models for the gravitational potential of the lens galaxy.
We review current estimates that the substructure represents 
$0.006 < f_{sat} < 0.07$ (90\% confidence) of the lens galaxy mass,
and outline future observational programs which can improve the results.  
\end{abstract}

\maketitle


\def\arcs{\ifmmode {^{\scriptscriptstyle\prime\prime}}
          \else $^{\scriptscriptstyle\prime\prime}$\fi}
\def\arcm{\ifmmode {^{\scriptscriptstyle\prime}}
          \else $^{\scriptscriptstyle\prime}$\fi}
\newdimen\sa  \newdimen\sb
\def\parcs{\sa=.07em \sb=.03em
     \ifmmode $\rlap{.}$^{\scriptscriptstyle\prime\kern -\sb\prime}$\kern -\sa$
     \else \rlap{.}$^{\scriptscriptstyle\prime\kern -\sb\prime}$\kern -\sa\fi}
\def\parcm{\sa=.08em \sb=.03em
     \ifmmode $\rlap{.}\kern\sa$^{\scriptscriptstyle\prime}$\kern-\sb$
     \else \rlap{.}\kern\sa$^{\scriptscriptstyle\prime}$\kern-\sb\fi}
\def\parcs{\hbox{$.\!\!^{\prime\prime}$}}
\def\arcs{\hbox{$^{\prime\prime}$}}
\def\gtorder{\mathrel{\raise.3ex\hbox{$>$}\mkern-14mu
             \lower0.6ex\hbox{$\sim$}}}
\def\ideq{\equiv}
\def\implies{\Rightarrow}
\def\inf{\infty}            
\def\ltlt{\ll}
\def\ltorder{\mathrel{\raise.3ex\hbox{$<$}\mkern-14mu
             \lower0.6ex\hbox{$\sim$}}}
\def\apj{ApJ}

\section{Introduction}

It is a generic feature of CDM (cold dark matter) halo simulations that
a significant fraction of the halo mass remains in the form of satellites
(e.g. Kauffmann et al.~1993, Moore et al.~1999,
Klypin et al.~1999).  The exact mass fraction
remains somewhat unclear, but the global mass fraction is of order
5--10\%, and the projected fraction inside cylinders of radius $R\sim
5h^{-1}$kpc is of order 1\%.  These mass fractions are significantly
higher than are 
observed in satellites of the Galaxy, suggesting a conflict
between CDM models and observations.  Three general classes of 
solutions have been proposed.  First, the satellites can be made
invisible by suppressing star formation 
(Kauffmann et al.~1993; Bullock et al.~2000).  
Second, they can be
destroyed by normal dynamical processes or abnormal ones such as
self-interacting dark matter (Spergel \& Steinhardt~2000;
Yoshida et al.~2000; Col{\'\i}n et al.~2002;
D'Onghia \& Burkert~2002).  Third, their formation might
be avoided by significantly reducing the amplitude of the power
spectrum on the relevant scales 
(Kamionkowski \& Liddle~2000; Col{\'{\i}}n et al.~2000;
Bode et al.~2001; Avila-Reese et al.~2001).  
The problem, of course,
is that it is difficult to distinguish between undetectable and
absent satellites.

\begin{figure}

  \includegraphics[height=.4\textheight]{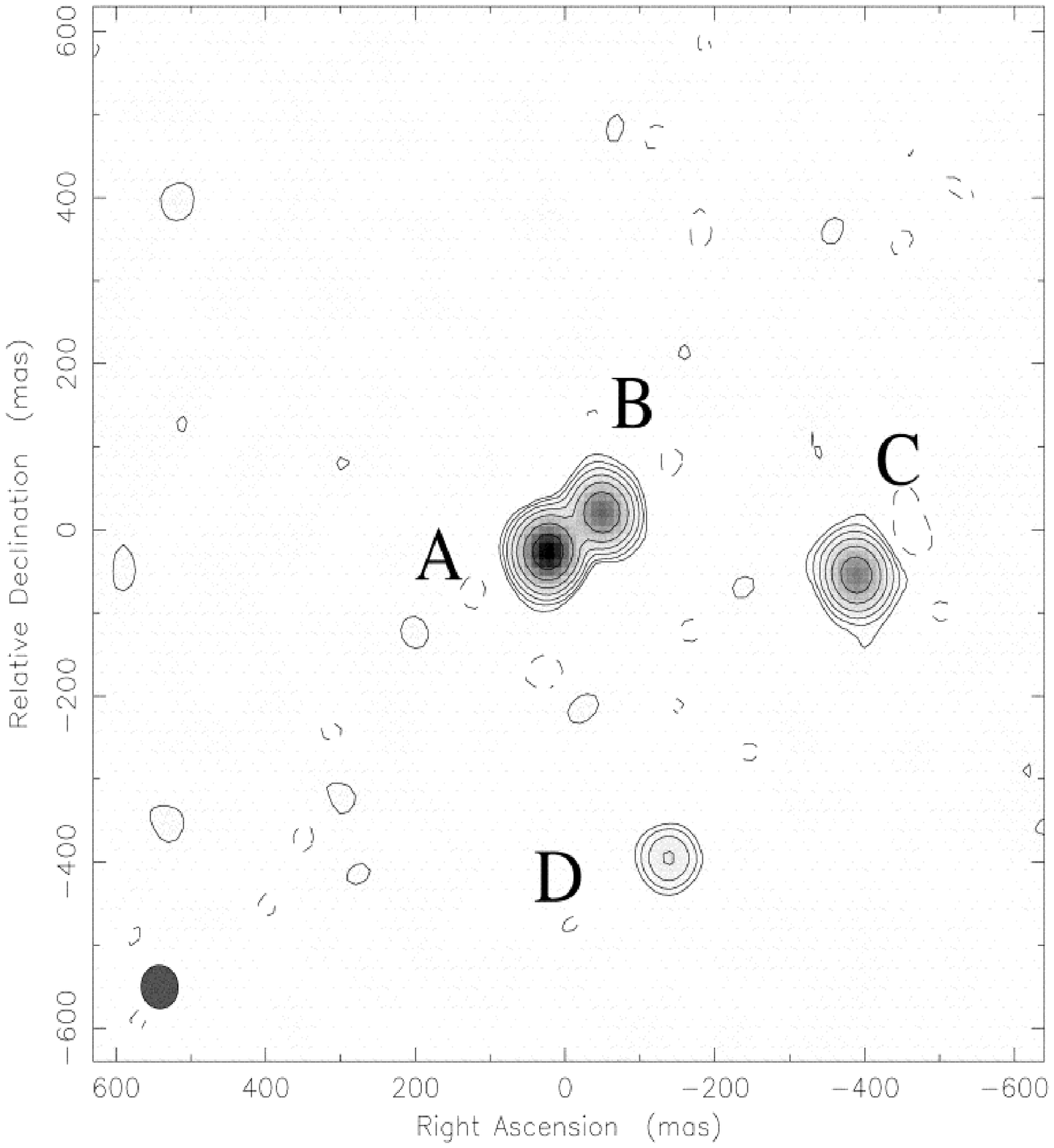}
  \includegraphics[height=.4\textheight]{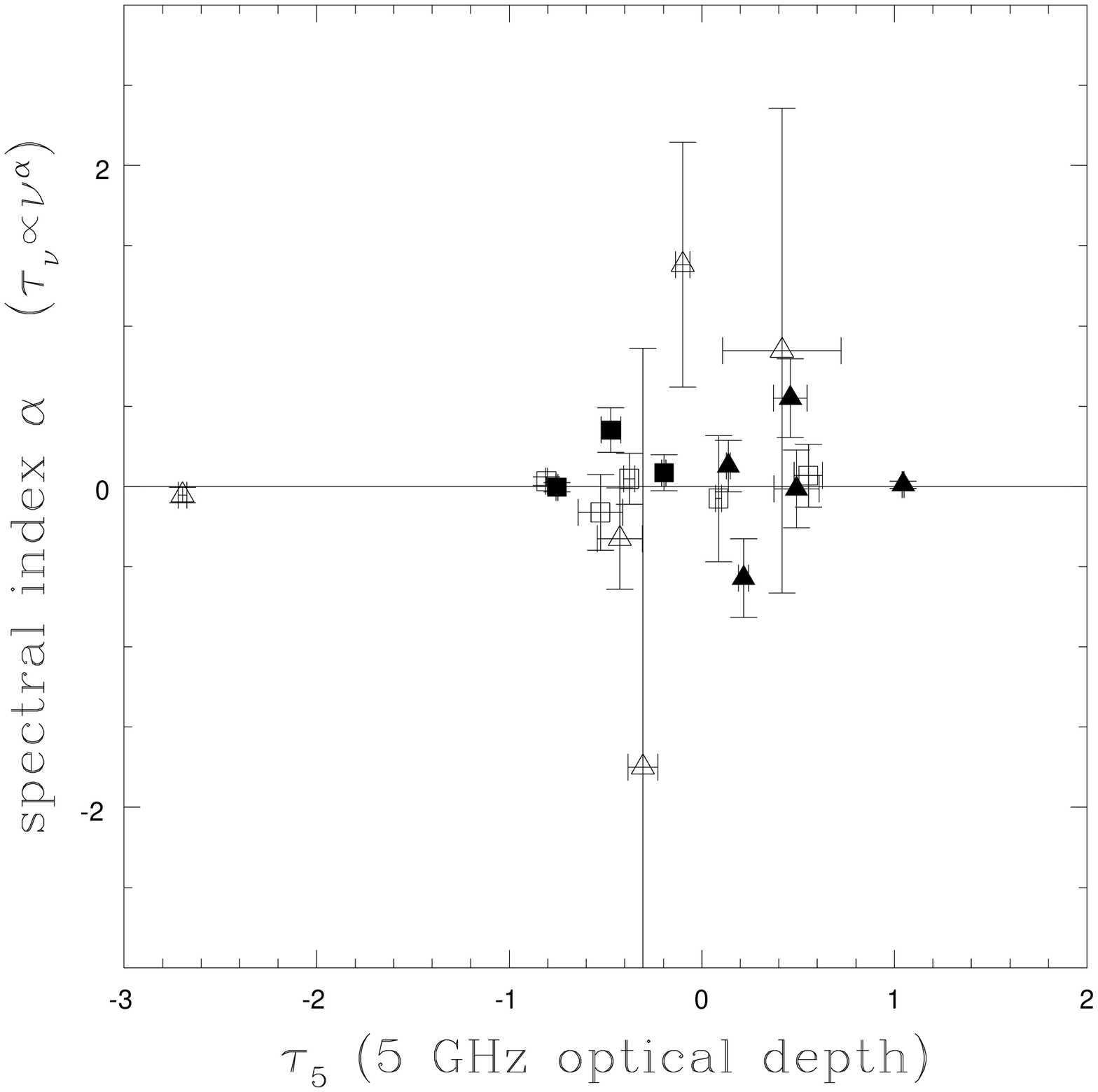}

  \caption{(LEFT) Example of an anomalous flux ratio.  For a smooth potential
  we would expect the A and B images in B1555+375 to have the same flux
  (Marlow et al.~1999).
  \newline
  (RIGHT) ISM properties needed to explain anomalous flux ratios.  For
  an optical depth function $\tau = \tau_5 (\nu/5\hbox{GHz})^\alpha$ we show
  the estimated spectral index $\alpha$ as a function of the optical
  depth $\tau_5$ at 5~GHz for the radio lenses used in DK02
  with published flux ratios at both 5~GHz and
  either 8 or 15~GHz. The points are coded by the image type:
  minima--squares, saddle points--triangles, brightest--filled, and
  faintest--open. }
  \label{fig:ism}
\end{figure}

It was realized early in the debate (see Moore et al.~1999)
that gravitational lensing provided a means of resolving the issue 
because it could detect the gravitational perturbations created by
substructure. It was already known that satellites provided a means
of solving the ``anomalous flux ratio'' problem seen in some 
gravitational lenses (Mao \& Schneider~1998).  An example
of such a problem is shown in Fig.~\ref{fig:ism}, where the close image pair
would be expected to have very similar fluxes for any lens model
where the gravitational potential can be well-represented by a 
low-order Taylor series expansion near the images.  
Metcalf \& Madau~(2001) 
and Chiba~(2002) pointed out that such anomalies should be
common given the predicted CDM substructure fractions. Mao \& 
Schneider~(1998), Keeton~(2001), 
Bradac et al.~(2002), and Chiba~(2002) explored
how substructure could explain the anomalous flux ratios in several
lens systems. Metcalf \& Zhao~(2001), Keeton, Gaudi 
\& Petters~(2002) and Evans \& Witt~(2002)
explored whether the model for the primary lens galaxy could be 
modified to explain the anomalous flux ratios, with mixed results
which we will discuss in detail below.  Our contribution in
Dalal \& Kochanek (2002, DK02 hereafter) was to analyze the data to
make an experimental determination of the substructure fraction,
finding it to be in the range
$0.006 < f_{sat} < 0.07$ (90\% confidence) based on a sample
of radio lenses.  This is in good agreement with the expectations
for CDM, and well above standard estimates for the mass fractions
in normal satellites.

Our review of the problem will cover three basic topics.  First, we
will discuss the problem of distinguishing CDM substructure from other
possible origins for the anomalous flux ratios.  In particular, we
introduce a simple statistical test for substructure in the gravity
as compared to either propagation effects in the interstellar medium
of the lens galaxy or poorly modeled contributions to the smooth gravitational
potential of the lens galaxy.  In \S3 we review our estimate of the
substructure mass fraction and its relation to simulations.  Finally,
in \S4 we discuss the future of the method. 

\section{Testing for Substructure}

Most of the existing studies of the anomalous flux ratio problem have 
focused on demonstrating that the problem can be explained by CDM 
substructure.  The next problem is to demonstrate that they cannot
be explained by other effects.  We can divide the other possibilities
into three categories.  First, propagation effects in the interstellar
medium of the lens galaxy could produce the observed anomalies.  Second,
the flux ratio anomalies could be created by problems in the models for
the smooth potential of the primary lens.  Third, we could be misinterpreting
a microlensing effect created by the normal stellar populations of the lenses
with the effects of more massive satellites.  Since we will focus on
radio lenses,  where microlensing effects must be small because of the
large source sizes (Koopmans \& de Bruyn~2000), we will not
discuss this effect in detail.

Here we explore the first two problems -- 
distinguishing substructure from propagation effects or modeling errors.
The first approach we could take is to argue individual cases.  For
example, almost all propagation effects should show a strong frequency
dependence.  One way to explore the required properties of the ISM
is to assume an optical depth, $\tau = \tau_5 (\nu/5\hbox{GHz})^\alpha$, 
for the radio lenses normalized by the optical depth $\tau_5$ at
$5$~GHz and with a spectral index $\alpha$ for the frequency dependence.
Fig.~\ref{fig:ism} shows the results of fitting such an ISM model to the radio
lenses in DK02 where flux measurements
were available at both 5~GHz and either 8 or 15~GHz.  Common ISM 
effects, such as refractive scattering or free-free absorption, would
show a spectral index of $\alpha \sim -2$, while the optical depth
function needed to explain the data has almost no frequency 
dependence ($\alpha \simeq 0$).  In short, explaining the anomalous
flux ratios with the ISM requires the radio equivalent of the 
``gray dust'' sometimes suggested to change the cosmological 
conclusions from Type Ia supernovae (Aguirre~1998).  

Similarly, Metcalf \& Zhao~(2001), Keeton, Gaudi
\& Petters~(2002) and Evans \& Witt~(2002)
explore whether changes to the smooth potential can explain the
problem.  The basic result from these studies is that they cannot.
Although Evans \& Witt~(2002) give an anti-substructure
tenor to their results, we would argue that they have actually
produced further arguments in favor of substructure.  First, for
the two lenses from DK02 they analyze,
they can only explain the anomalous flux ratio of one system 
despite using lens models with essentially arbitrary angular
structure.  In fact, when we use similar models to analyze the
full sample from DK02, we find that
of the 6 systems (out of 7) which arguably have anomalous flux
ratios, the more complicated models can successfully fit 2--3, at the
price of having amplitudes for the higher order perturbations that
are significantly larger than are generally observed for either the
stars or in dark matter simulations.
Second, the other two lenses Evans \& Witt~(2002) analyze are known
from time variability studies to be microlensed (i.e. containing
substructure but on a smaller mass scale), so the success of the
Evans \& Witt~(2002) models at explaining the flux ratios
in these systems shows that sufficiently complex macro models can
mask the presence of substructure even when it is known to be 
present.  This leaves us with a basic ambiguity of course, since
standard models (ellipsoidal lens models combined with external
tidal shear fields) cannot explain the anomalous flux ratios, 
while models with very complicated potentials can explain some,
but not all, anomalous flux ratios, but can also do so in systems
where they should not.

Fortunately, we need not live with these ambiguities, because low
optical depth substructure has a unique property that
allows us to statistically distinguish substructure from either
the interstellar medium or problems in the smooth potential.
The images of a lens can be assigned a parity depending on 
whether they are saddle points or minima of the virtual time
delay surface (e.g. Schneider et al.~1992), 
and the four images alternate their parities 
as we go around the Einstein ring (saddle-minimum-saddle-minimum).
Given a sample of lenses, we can divide the images into four 
separate classes (brightest saddle, faintest saddle, brightest
minimum, faintest minimum) based on their parities and fluxes.
As we now discuss, the effects of substructure depend on the image
type, while the effects of the ISM and errors in the macro model
generally do not.

\begin{figure}

  \includegraphics[height=.6\textheight]{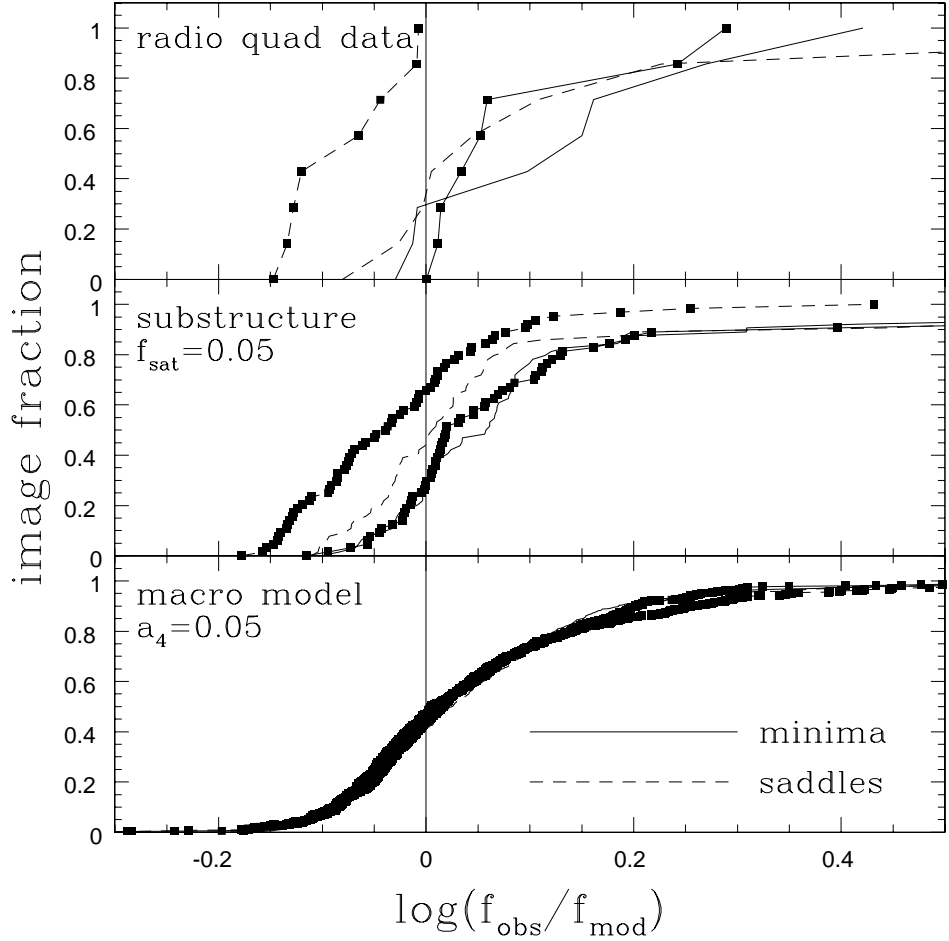}

  \caption{Residual distributions after fitting standard models.  
    Each panel shows the integrated distributions of the flux
    residuals $\log(f_{obs}/f_{mod})$ given the observed $f_{obs}$
    and best fitting model $f_{mod}$ image fluxes.  The distributions
    are shown separately for the brightest (points) and faintest
    (no points) minima (solid lines) and saddle points (dashed lines).
    The top panel shows the distribution for the real data on 8
    four-image radio lenses.  The middle panel shows a Monte Carlo
    simulation of the distributions for lenses with a $f_{sat}=0.05$
    mass fraction in substructure modeled as tidally truncated 
    singular isothermal spheres.  The lower panel shows a Monte
    Carlo simulation of the distributions for lenses with a large
    amplitude, randomly oriented $\cos 4\theta$ term in the 
    gravitational potential of the lens that is not included
    in the lens model used to interpret the data.  The mode amplitude of
    $a_4=0.05$ is the standard Fourier component used to analyze
    the photometry of elliptical galaxies.  Note how the data (top) shows
    the same shift to fainter fluxes for the brightest saddle points
    as is expected from low optical depth substructure (middle), while an
    error in the macro model (bottom) does not distinguish between the image
    types.
    }
  \label{fig:statistics}
\end{figure}

Both microlensing by the stars (Schechter \& Wambsganss 2002)
and lensing by extended substructures (Keeton 2002) distinguishes
between images based on their parities when the optical depth
is low.  The sense of the effect is to preferentially demagnify
saddle points (negative total parity) compared to minima (positive
total parity).  The most magnified images are also affected more
than the least magnified images because the high magnification makes
them sensitive to smaller perturbations in the potential (Mao \& 
Schneider~1998).  We can study this effect by examining
the distributions of the residuals, $\log(f_{obs}/f_{mod})$, between the model 
fluxes $f_{mod}$ and observed fluxes $f_{obs}$ for both the data
and for different theories as to the origin of the anomalous flux
ratios for the 4 different image types found in a quad lens
(brightest saddle, faintest saddle, brightest minimum, faintest
minimum).  The top panel of Fig.~\ref{fig:statistics} shows the distributions 
we find after fitting the 8 available radio quads using our standard model
for the smooth potential (one or more singular isothermal ellipsoids
in an external tidal shear field), and the middle panel shows the
distribution predicted in a Monte Carlo simulation of a lens sample
with a 5\% substructure mass fraction.  As expected from the previous
theoretical studies, the simulation shows an offset of the distribution
for the brightest saddle points from the distribution for the other
images, in the sense of preferentially demagnifying the saddle point.

The ISM, for example, makes a very different prediction.
It is the clumpy, high density 
components of the ISM which will modify flux ratios, and this
has two important consequences.  First, propagation effects 
should preferentially modify the fluxes of the {\it least} magnified 
images, because they have the smallest 
intrinsic angular sizes.  The more magnified images have larger
angular sizes and will more effectively smooth out any effects of 
a clumpy ISM.    Second, a clumpy ISM cannot distinguish between images 
of differing parities because the ISM properties are locally determined
while the image parity is not -- the magnification tensor depends on the
projected surface density, the projected shear component of the gravity
and the source and lens redshifts.  Thus, in a statistical sample, 
the ISM might systematically
perturb the fainter images more than the brighter images, but it will
not distinguish between saddle points and minima.

The macro model also will have difficulty systematically perturbing
images of a particular parity.  Qualitatively this can be understood
by the symmetry of merging image pairs from the point of view of the
central potential -- any slope in the curvature needed to produce 
a change in the magnification of the saddle point can just as easily
appear with the opposite sign so as to produce the opposite change.
While we lack a mathematical proof to this effect, it certainly holds
in Monte Carlo simulations of lenses produced by potentials with
complicated, higher order angular structures (as in Evans \& Witt~2002)
that are then modeled using standard ellipsoidal potentials.  In
Fig.~\ref{fig:statistics} we show an example of the distribution of residuals 
expected for a population of lenses with a large, unmodeled $\cos 4\theta$ 
perturbation to the potential. Unlike the distributions expected
for substructure shown in Fig.~\ref{fig:statistics}, we see that there is no 
distinction in the model residuals for the different images.

This then leads to a simple test for substructure -- do the distributions
of residuals from standard ellipsoidal models distinguish between the
image types or do they not?  We can immediately see the
answer in Fig.~\ref{fig:statistics} -- the distribution of residuals for the 
brightest saddle point differs from that of the other three images just as 
expected for substructure.  Quantitatively, the Kolmogorov-Smirnov test 
probability that the model residuals for the brightest saddle points have 
the same distribution as for the other images is $<0.1\%$.  
Other phrasings of the test, 
bootstrap resampling of the data, null tests with the image identifications
randomly assigned all support the results that the statistical properties
of the saddle points are fundamentally different from that of the other
images.  And this distinctiveness is the tell-tale sign that the anomalous
flux ratios are due to substructure in the gravity rather than the ISM
or problems in the smooth potential model for the lens galaxy.

\section{Implications for CDM }

If substructure is the explanation, then the next objective is to estimate
the satellite mass fraction $f_{sat}$.  Here we review the method and results
of DK02.  We modeled the substructure as
tidally truncated singular isothermal spheres, with a critical radius scale
$b=0\parcs001$ and a tidal radius $a=(b b_0)^{1/2}$ where $b_0\simeq 1\parcs0$ 
is the critical radius of the primary lens galaxy.  When the dominant
effect of the substructure is to perturb magnifications rather than
image positions, we can only measure the mass fraction $f_{sat}$ of the satellites
with reasonable accuracy.  The mass scale or mass function of the satellites
is difficult to constrain unless there are significant astrometric perturbations.
In outline, our approach was to take each lens and its standard model and then
add random substructure realizations in order to determine the probability of
finding an improved fit as a function of the parameters describing the
substructure (principally the mass fraction, $f_{sat}$).  Because the 
``macro'' model for the smooth potential masks some of the effects of substructure,
it is necessary to reoptimize the parameters of the ``macro'' model for
every trial.  We applied the method to a sample of 7 four-image lenses.

\begin{figure}

  \includegraphics[height=.34\textheight]{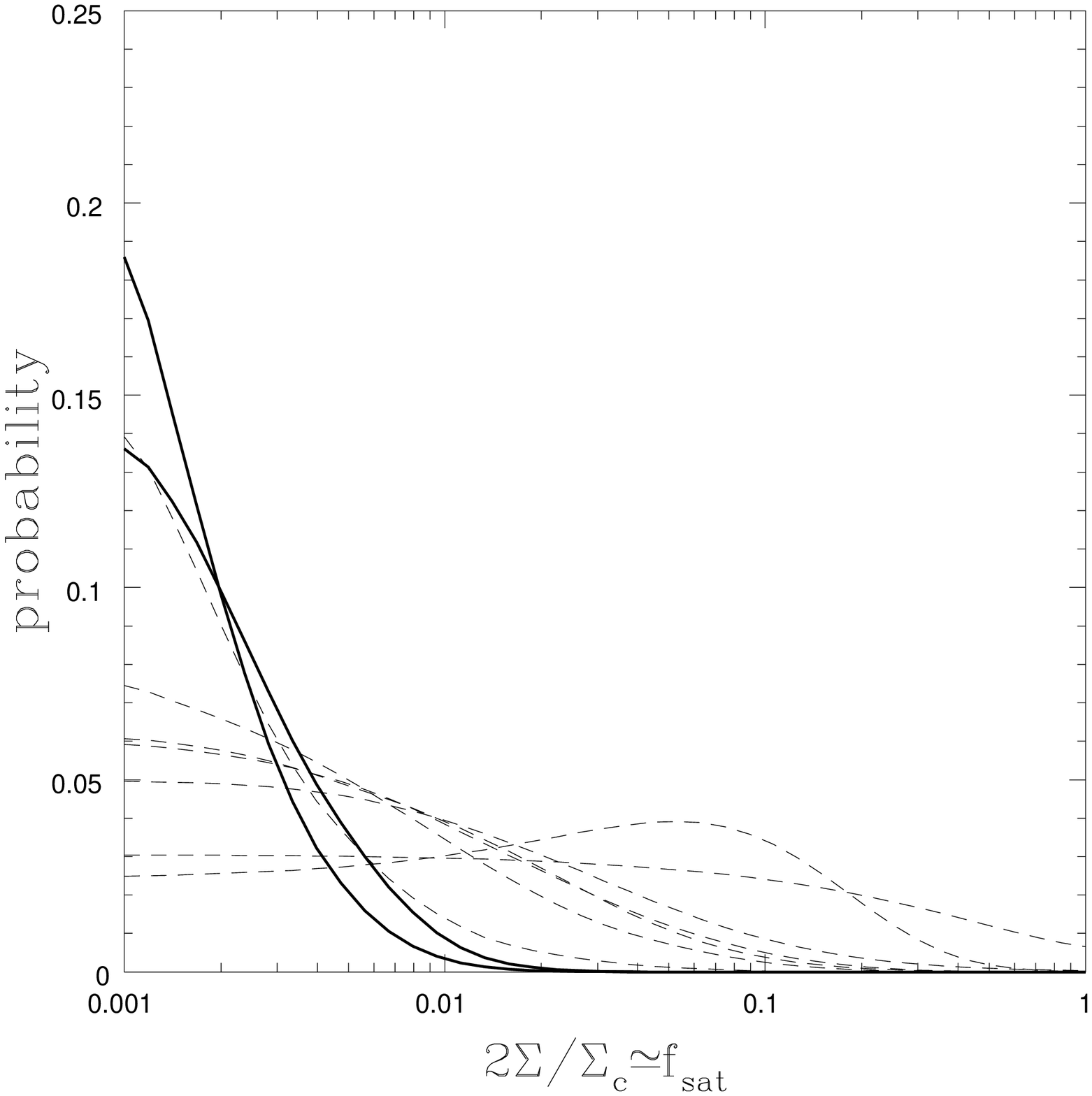}
  \includegraphics[height=.34\textheight]{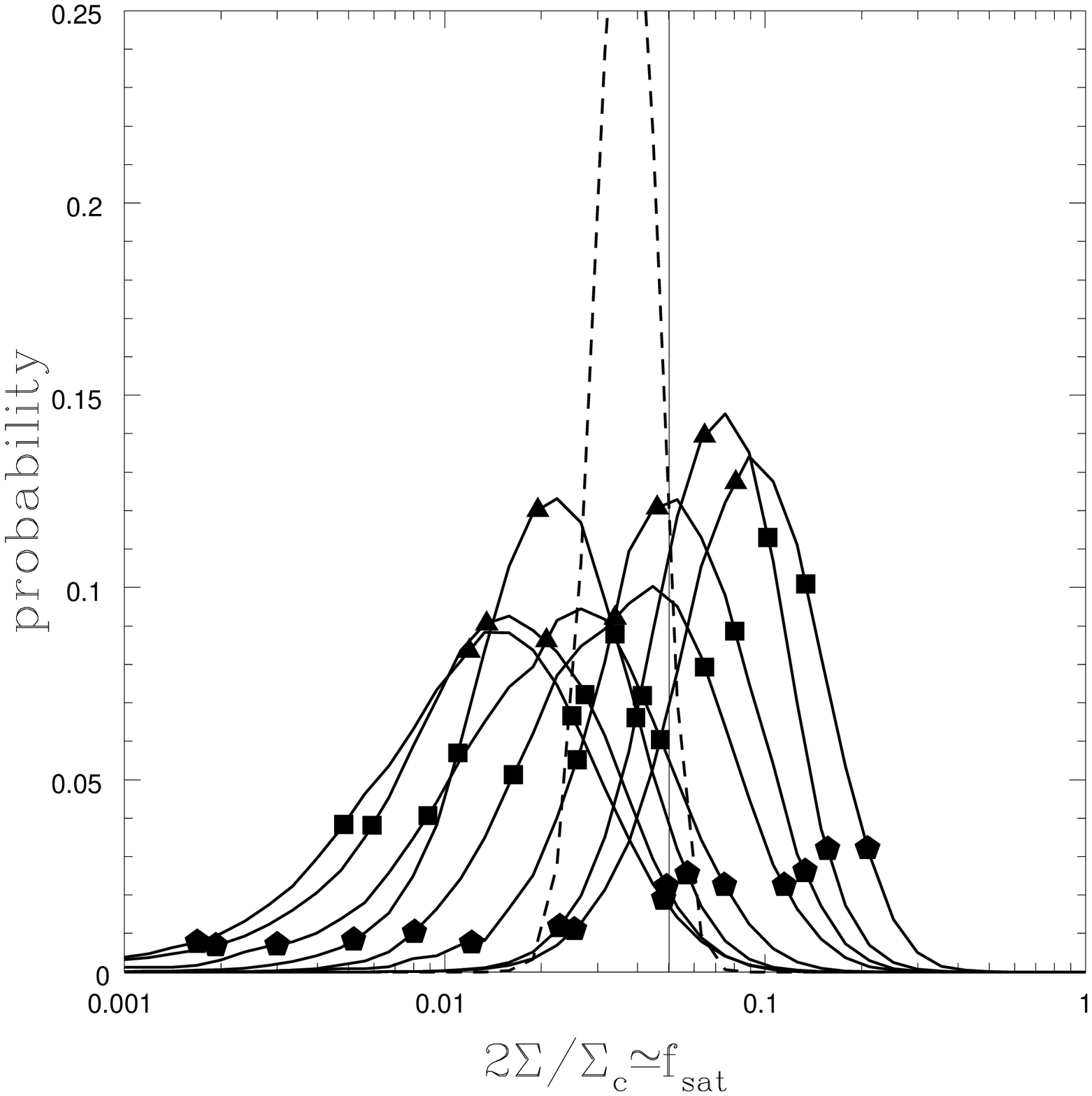}

  \caption{Monte Carlo simulations of the DK02 method.  The left (right) panel 
  shows the likelihood distributions for the substructure mass fraction 
  produced by our method when $f_{sat}=0$ ($f_{sat}=0.05$).  In the left
  panel, the dashed lines show the probabilities for the individual lenses
  and the solid lines show the joint likelihoods for two simulations of a
  sample of 7 lenses.  For these simulations where the true $f_{sat}=0$,
  we find an upper bound of $f_{sat}\ltorder 0.004$.  In the right panel
  we show the joint likelihoods for eight simulations of a sample of 7
  lenses (solid lines) in which the true substructure fraction is $f_{sat}=5\%$.  
  The recovered satellite fractions are statistically consistent with the 
  input fraction, albeit with broad uncertainties due to the small sample
  size.  The heavy dashed line simulates a sample of 56 lenses (the product
  of all the solid lines), and the recovered value is slightly lower than
  the input value given the distribution width.  The points on the curves
  indicate the median (triangles), 68\% confidence (squares) and 95\%
  confidence (pentagons) regions.
    }
  \label{fig:monte}
\end{figure}

We can illustrate our method with Monte Carlo simulations. Fig.~\ref{fig:monte} 
shows the results for Monte Carlo simulations of our sample either with or without
substructure.  If we add no substructure, then we typically obtain an 
upper limit of $f_{sat}\ltorder 0.004$ given the properties of our 
lens sample.  This sets a lower limit for our detection threshold somewhat
above the substructure fraction which would be associated with the visible
satellites in the Galaxy.  If we put $f_{sat}=0.05$ of the mass into
substructure, then we recover the input fraction reasonably accurately.
Of the eight Monte Carlo trials shown in Fig.~\ref{fig:monte}, 
four agree with the
input value to within the 68\% ($1\sigma$) confidence region, and six
agree to within the 90\% confidence region.  If we combine all 8 
simulations into a synthetic sample of 56 lenses, we estimate that
$f_{sat}=0.034$ with a 90\% confidence range of $0.023 \leq f_{sat} \leq 0.048$
that marginally excludes the true value.  Similarly, if we attempt to
recover the deflection scale of the substructure, the error bars are
worse but the results do converge to the input value when we model a 
large enough sample of lenses.

\begin{figure}

  \includegraphics[height=.34\textheight]{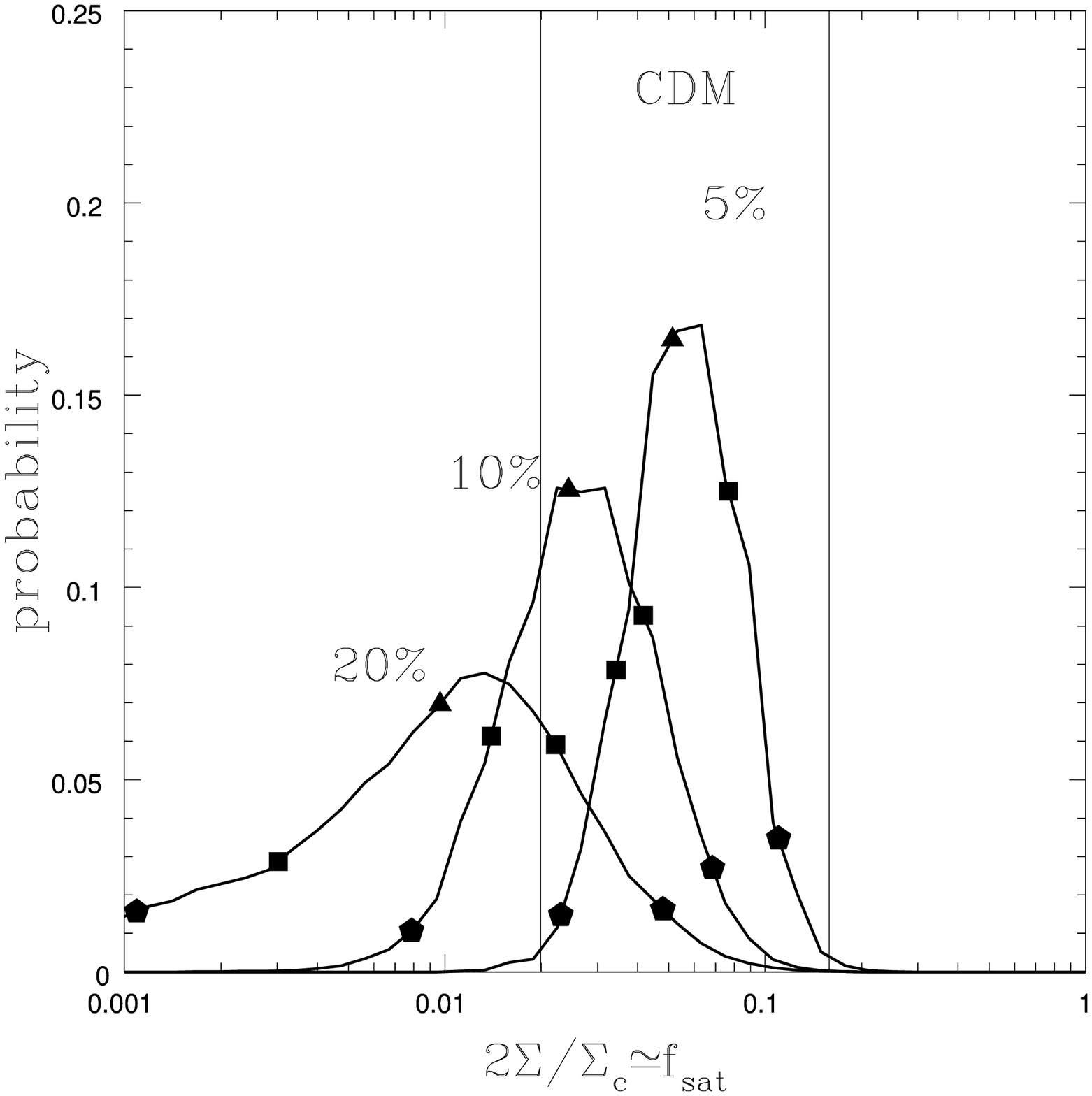}
  \includegraphics[height=.34\textheight]{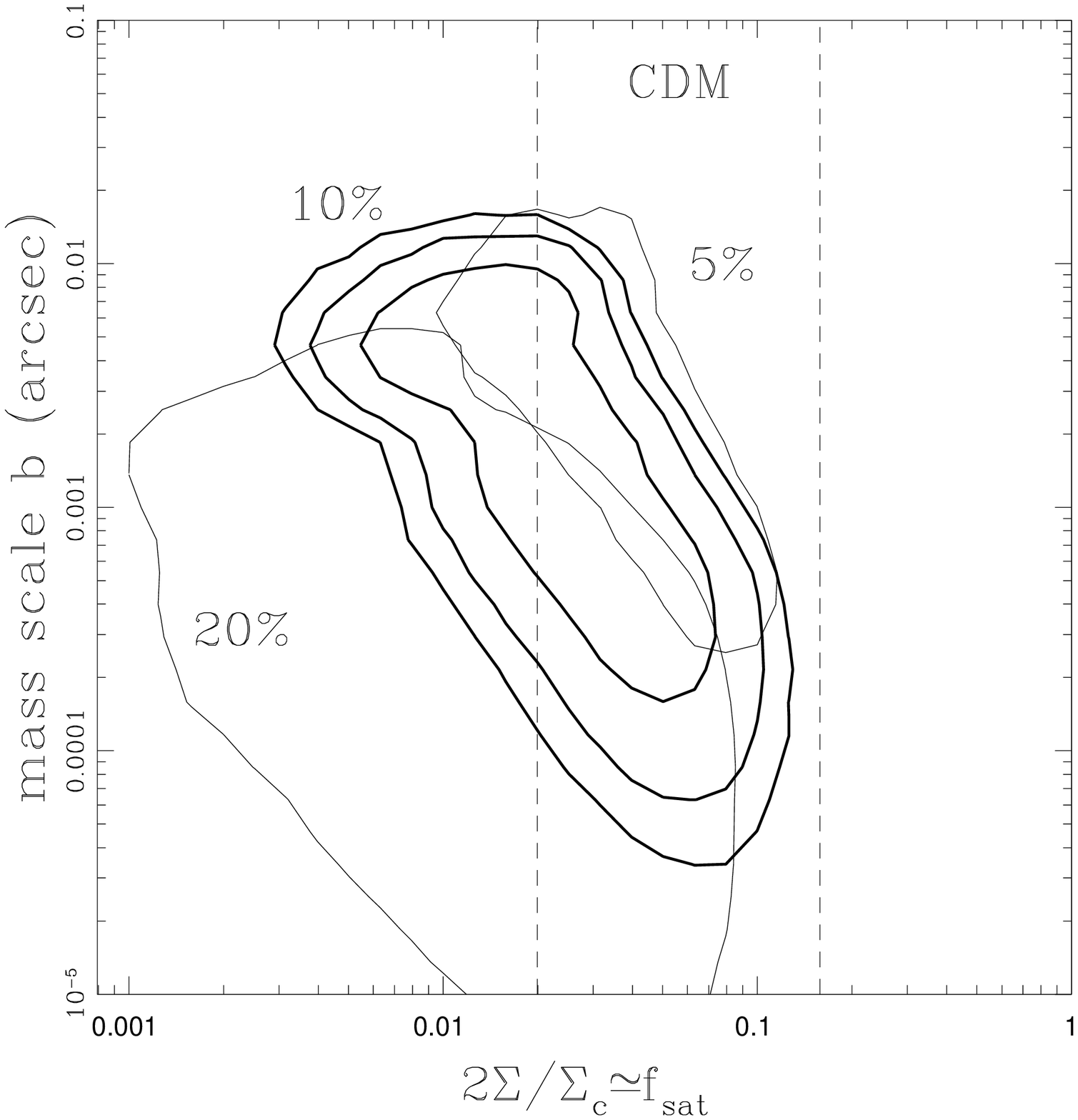}

  \caption{Results for the DK02 sample of 7 lenses.  
  The left panel shows the results for estimating
  $f_{sat}$ with $b=0\protect\parcs001$ fixed, and the right panel shows the
  results estimating the deflection scale $b$ as well.  Distributions are
  shown assuming flux measurement errors of 5\%, 10\% and 20\%, where
  we adopted 10\% as our standard estimate.  
    }
  \label{fig:data}
\end{figure}

Fig.~\ref{fig:data} shows the results for the real data.  The biggest systematic 
uncertainty
in the data is the level of systematic uncertainty in the flux ratio measurements,
so we show the results for 5\%, 10\% and 20\% uncertainties in flux measurements.
We adopt 10\% as our standard value -- the measurements probably are not as
accurate as 5\%, they probably are more accurate than 10\%, and they certainly 
are more accurate than 20\%.  With the
10\% flux uncertainties we find a substructure fraction of 
$0.006 \ltorder f_{sat} \ltorder 0.07$ (90\% confidence) that is in good agreement
with the expectations for CDM models.  We obtain a very poor estimate of
the characteristic deflection scale, finding $0\parcs0001 < b < 0\parcs007$,
which for a substructure mass function $dn/dM \propto M^{-1.7}$ implies an
upper end to the substructure mass function of $10^6M_\odot$-$10^9 M_\odot$ 
that is in crude accord with our expectations.  The degenerate direction in
the error contours of Fig.~\ref{fig:data} correspond to keeping the magnification 
perturbations nearly constant while varying the astrometric perturbations.

\section{In The Future}

The future of the substructure question will be driven by further observations,
both to clarify the origins of the anomalous flux ratios and to obtain 
improved estimates of the substructure mass fraction and mass function.

It is relatively straight forward to finish eliminating the ISM as a source of 
concern with new observations.  In the radio this means measuring flux ratios at 
still higher frequencies (e.g. 43~GHz flux ratio measurements at the VLA)
to constrain the frequency dependence of any propagation effect still more
tightly.  In the optical this means measuring flux ratios over long 
wavelength baselines to measure any dust extinction 
(e.g. Falco et al.~1999).  Mid-infrared (5--10$\mu$m) flux 
ratios, where the wavelength is far to short to be bothered by electrons and 
far too long to be bothered by dust, are difficult to measure but completely
insensitive to the ISM.
Observations to find additional lensed structures in the systems with 
anomalous flux ratios are the best direct route to determining whether
more complicated lens potentials are needed.  In particular, very clean
constraints on the strengths of any more complicated angular structure
than is included in the standard ellipsoidal models can be obtained by
analyzing the shapes of the Einstein ring images of the host galaxies
(see Kochanek et al.~2001).  Such data can be obtained for
any lens through deep HST/NICMOS imaging of lens systems.  

Improving estimates of the substructure parameters or the statistical case
for (or against) substructure requires larger samples of lenses to include
in the analysis.  The primary problem in expanding the sample is the need
to separate the effects of stars and satellites in the optically-selected
lenses.   The flux ratios of the optical quasars are affected by
both substructure and stellar microlensing because the optical continuum
emitting regions of accretion disks are so compact (see
Schneider et al.~(1992) for a general review of microlensing).
By measuring the flux ratios of these lenses in either the mid-IR, where the
emitting region is a large dust ``torus,'' or in the emission lines, where the
emitting region is the relatively large broad/narrow emission line region,
we can separate the effects of the stars from the effects of
substructure (e.g. Moustakas \& Metcalf~2002).  While mid-IR
imaging is difficult except for the brightest quasar lenses 
(see Agol et al.~2000), the advent of high spatial resolution 
integral field spectrographs on many 8m-class telescopes will make it
relatively easy to measure emission line flux ratios. 

\begin{figure}

  \includegraphics[height=.3\textheight]{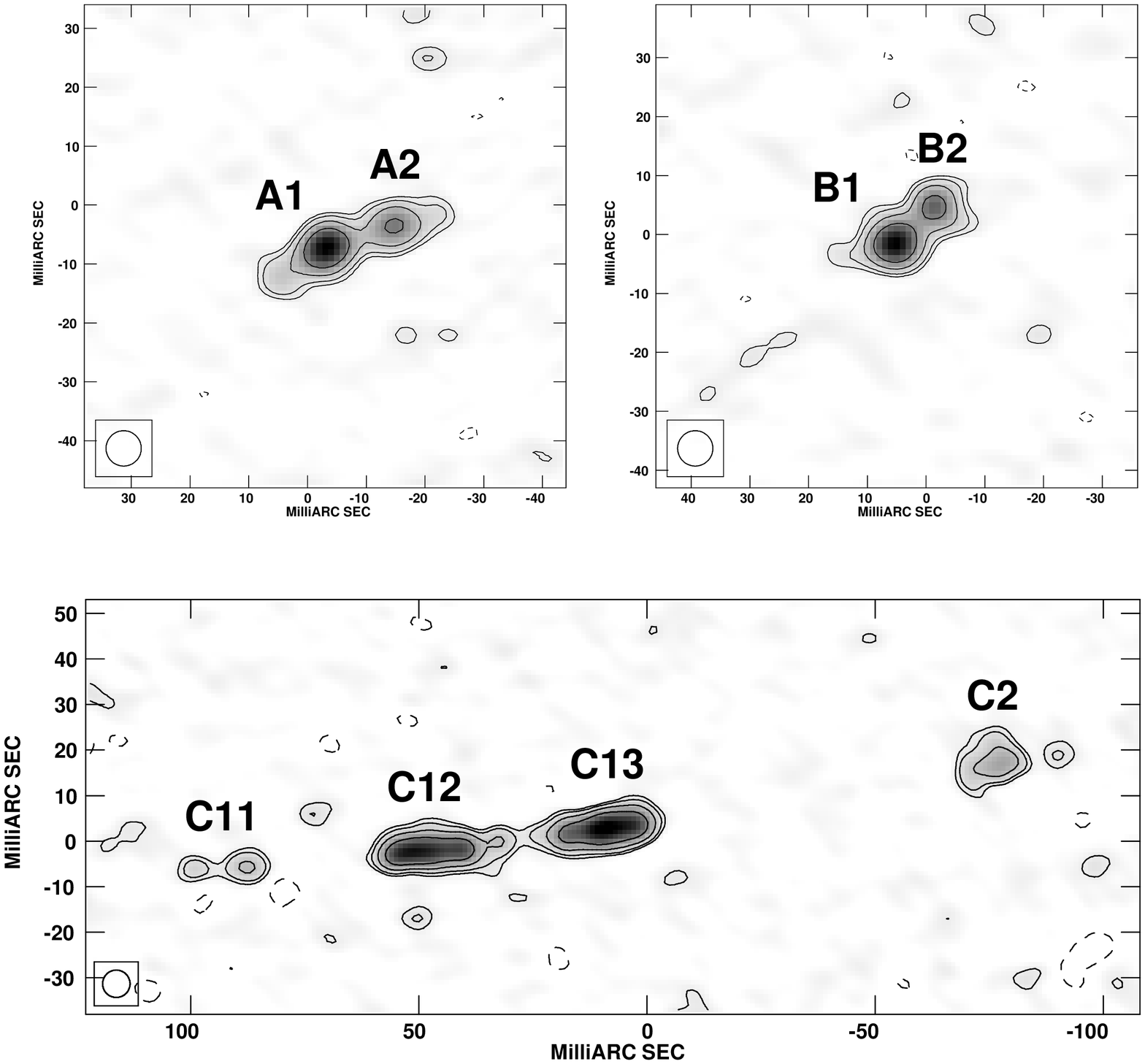}
  \includegraphics[height=.3\textheight]{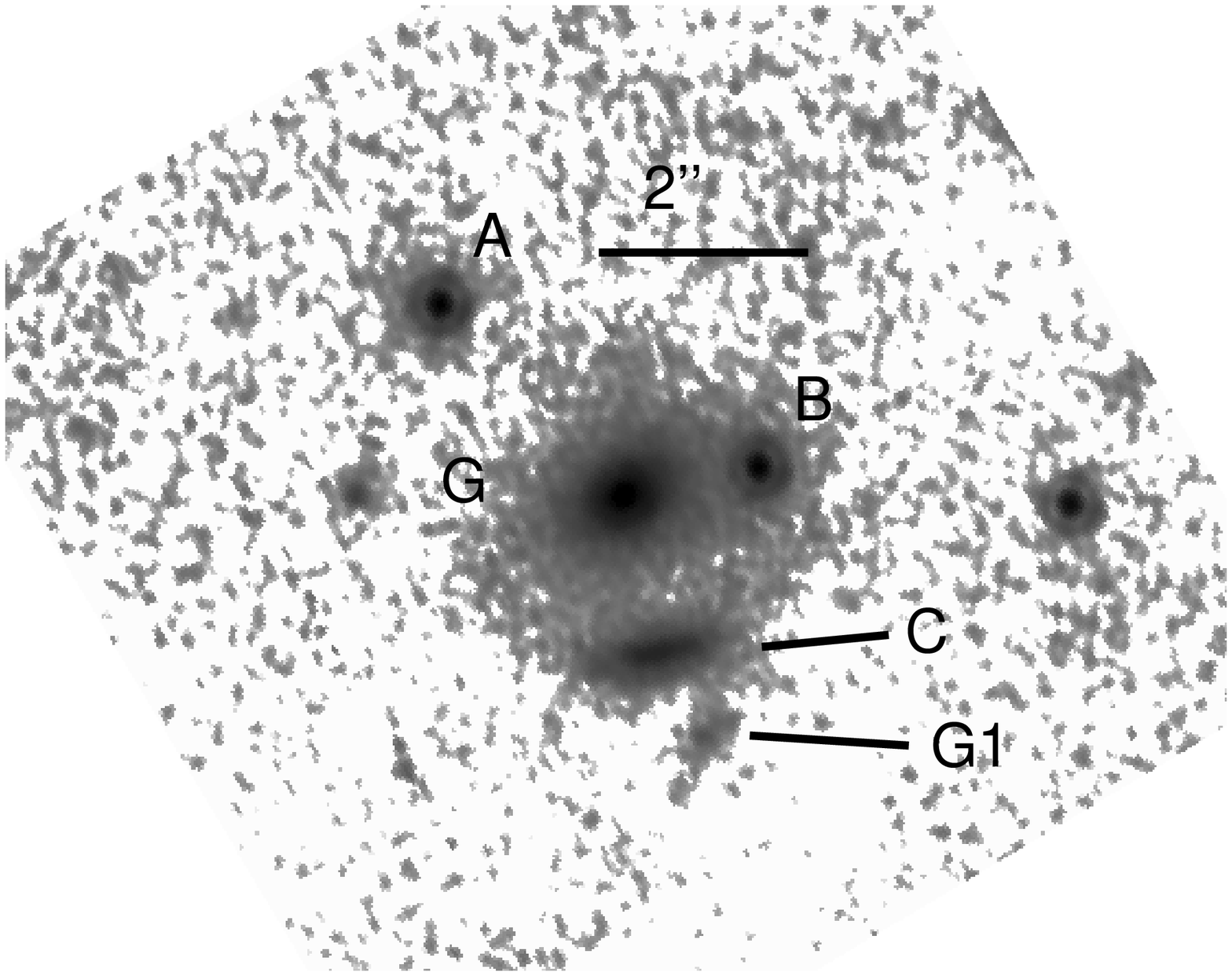}

  \caption{The Astrometric anomaly in MG2016+112.  The left panels show the VLBI
  images from Koopmans et al.~(2001) of the A, B and
  C images.  The right panel shows the CASTLES HST/NICMOS H-band image of the
  system.  For the same symmetry reasons that close image pairs should have
  the same fluxes, the C$_{11}$--C$_{12}$ and C$_{13}$--C$_2$ image separations
  should be the same.  The fact that they differ is the astrometric equivalent
  of a flux anomaly.  In this case it is created by the small galaxy G1
  sitting to the South of the C images.
   } 
  \label{fig:2016}
\end{figure}

Our analysis methods also need to be improved.  In particular, we need to properly
treat the highest mass satellites, both to constrain
the satellite mass function and to better estimate the mass fraction.  In the
DK02 analyses, we included the highest mass satellites as part of the macro
model because they have such an enormous effect on the models that it is 
impossible to produce a reasonable model without including them.  One example
is the small satellite in MG0414+0534 (Object ``X''; 
Schechter \& Moore 1993), which at H-band
has only 10\% the luminosity of the main lens and in models has only 
12\% the critical radius of the main lens, but even when fitting only the
positions of the quasar images produces a $\Delta\chi^2 \simeq 100$ 
improvement in the fit once it is included (Ros et al.~2000).  

We can also search for the
astrometric equivalents of anomalous flux ratios so as to provide better
constraints on the mass scale of the substructure (e.g Wambsganss \&
Paczynski~1992, Metcalf~2002).
One example is the lens MG2016+112 (see Fig.~\ref{fig:2016}) where in 
VLBI maps the  C image is seen to be a pair of merging images
each of which is composed of two VLBI components 
(Koopmans et al.~2001).  For the same
reasons that we would expect a merging image pair to have similar
fluxes, we would expect them to have similar separations, so the 
very asymmetric separations of the VLBI components C$_{11}$--C$_{12}$ 
as compared to C$_{13}$--C$_2$ is the astrometric equivalent of an
anomalous flux ratio.  In this case, as in MG0414+0534, the culprit
is a visible satellite G1 sitting to the south of the C image complex. 
It has only 1-2\% the H-band luminosity and $\simeq 8\%$ the deflection scale
of the main lens, and is known to lie at the same redshift as the lens
galaxy.  The VLBI data has sufficient resolution to detect a satellite
with a deflection scale nearly 10 times smaller.  The holy grail of
searching for CDM substructure in gravitational lenses would be to 
find an astrometric anomaly similar to that in MG2016+112, so that
there is compelling evidence for the existence of a satellite, but for which
no luminous counterpart can be detected.


\begin{theacknowledgments}
N.D. gratefully
acknowledges the support of NASA through Hubble Fellowship grant
\#HST-HF-01148.01-A awarded by STScI, which is operated by AURA
for NASA, under contract NAS 5-26555.
CSK is supported by the Smithsonian Institution and NASA grant NAG5-9265.
\end{theacknowledgments}

\bibliographystyle{aipproc}

\end{document}